\documentstyle[12pt,amsfonts]{article}
 
\makeatletter
\@addtoreset{equation}{section}
  \def\theequation{\thesection.\arabic{equation}}
\def\eqnarray{\stepcounter{equation}\let\@currentlabel=\theequation
\global\@eqnswtrue
\global\@eqcnt\z@\tabskip\@centering\let\\=\@eqncr
$$\halign to \displaywidth\bgroup\@eqnsel\hskip\@centering
  $\displaystyle\tabskip\z@{##}$&\global\@eqcnt\@ne 
  \hfil$\displaystyle{\hbox{}##\hbox{}}$\hfil
  &\global\@eqcnt\tw@ $\displaystyle\tabskip\z@
  {##}$\hfil\tabskip\@centering&\llap{##}\tabskip\z@\cr}
\def\lefteqn#1{\hbox to 2em{$\displaystyle #1$\hss}}
\mathchardef\by="202
\mathchardef\Gamma="100
\def\cn{\mathop{\rm cn}\nolimits}
\def\dn{\mathop{\rm dn}\nolimits}
\def\sn{\mathop{\rm sn}\nolimits}
\def\f #1#2{\mathop{f_{#1}}\limits^{(#2)}{}}
\def\g #1{\mathop{g}\limits^{(#1)}{}}
\makeatother

\begin{document}

\title{Riemannian Manifolds with Diagonal Metric.\\
The Lam\'e and Bourlet Systems} 

\author{Alexander V. Razumov\thanks{E-mail: razumov@mx.ihep.su}\hskip
0.8em and Mikhail V. Saveliev\thanks{E-mail: saveliev@mx.ihep.su} \\ 
{\small \it Institute for High Energy Physics,
142284, Protvino, Moscow region, Russia}}

\date{}

\maketitle

\begin{abstract}
We discuss a Lie algebraic and differential geometry construction of
solutions to some multidimensional nonlinear integrable systems
describing diagonal metrics on Riemannian manifolds, in particular
those of zero and constant curvature. Here some special solutions to
the Lam\'e and Bourlet type equations, determining by $n$ arbitrary
functions of one variable are obtained in an explicit form. For the
case when the sum of the diagonal elements of the metric is a
constant, these solutions are expressed as a product of the Jacobi
elliptic functions and are determined by $2n$ arbitrary constants.
\end{abstract}

\section{Introduction}

The classical differential geometry serves as an injector of many
equations integrable in this or that sense. Among them, the Lam\'e
and Bourlet equations play especially remarkable role. These
equations arise, in particular, in the following way.

Let $(U; z^1, \ldots, z^n)$ be a chart on a Riemannian manifold
$(M, g)$, such that the metric tensor $g$ has on $U$ the form
\begin{equation}
g|_U = \sum_{i=1}^n \beta_i^2 \, dz^i \otimes dz^i. \label{1.4}
\end{equation}
In such a situation the metric tensor $g$ is said to be {\it
diagonal} with respect to the coordinates $z^i$.  The functions
$\beta_i$ are called the {\it Lam\'e coefficients}. Introduce an
orthonormal basis in the space of 1-forms on $U$ defining
\[
\theta^{\hat \imath} = \beta_i dz^i.
\]
Here and in what follows we use for the indices referring to the
orthonormal basis the same letters as for ones referring to the
coordinate basis, but supply them with a hat. Note that for an
orthonormal basis there is no actual distinction between lower and
upper indices. In terms of the forms $\theta^{\hat \imath}$ the
metric tensor is written as 
\[
g|_U = \sum_{i=1}^n \theta^{\hat \imath} \otimes \theta^{\hat
\imath}. 
\]

Let us find the curvature two-forms $\Omega^{\hat \imath}{}_{\hat
\jmath}$ of the Levi-Civita connection corresponding to the
metrics given by (\ref{1.4}). The simplest way here is to use the
second Cartan structural equation \cite{KNo63}
\begin{equation}
\Omega^{\hat \imath}{}_{\hat \jmath} = d \omega^{\hat \imath}{}_{\hat
\jmath} + \sum_{k=1}^n \omega^{\hat \imath}{}_{\hat k} \wedge
\omega^{\hat k}{}_{\hat \jmath}, \label{1.8}
\end{equation}
where $\omega^{\hat \imath}{}_{\hat \jmath}$ are connection 1-forms
related to the connection coefficients $\Gamma^{\hat \imath}{}_{\hat
\jmath \hat k}$ by the equality 
\[
\omega^{\hat \imath}{}_{\hat \jmath} = \sum_{k=1}^n \Gamma^{\hat
\imath}{}_{\hat \jmath \hat k} \theta^{\hat k}.
\]
For the connection coefficients in the case of an orthonormal basis
one has the expression
\[
\Gamma^{\hat \imath}{}_{\hat \jmath \hat k} = \frac{1}{2} ( c^{\hat
k}{}_{\hat \imath \hat \jmath} + c^{\hat \jmath}{}_{\hat \imath \hat
k} - c^{\hat \imath}{}_{\hat \jmath \hat k}),
\]
where 
the functions $c^{\hat \imath}{}_{\hat \jmath \hat k}$ are determined
by the relation 
\[
d \theta^{\hat \imath} = - \frac{1}{2} \sum_{i,j=1}^n c^{\hat
\imath}{}_{\hat \jmath \hat k} \theta^{\hat \jmath} \wedge
\theta^{\hat k}.
\]
Defining the so called {\it rotation coefficients}
\begin{equation}
\gamma_{ij} = \frac{1}{\beta_i} \partial_i \beta_j, \qquad i \ne j,
\label{1.3} 
\end{equation}
we find 
\[
d \theta^{\hat \imath} = \frac{1}{\beta_i} \sum_{j=1}^n \gamma_{ji}
\theta^{\hat \jmath} \wedge \theta^{\hat \imath}.
\]
{}From this relation it follows that
\[
c^{\hat \imath}{}_{\hat \jmath \hat k} = \frac{1}{\beta_i}
(\gamma_{ki} \delta_{ij} - \gamma_{ji} \delta_{ik}),
\]
therefore, in the case under consideration one has 
\[
\Gamma^{\hat \imath}{}_{\hat \jmath \hat k} = \frac{1}{\beta_k}
(\gamma_{jk} \delta_{ki} - \gamma_{ik} \delta_{kj}),
\]
and we come to the following expression for the connection forms
\[
\omega^{\hat \imath}{}_{\hat \jmath} = \gamma_{ji} dz^i - \gamma_{ij}
dz^j.
\]
Substituting this expression in (\ref{1.8}), we obtain
\begin{eqnarray}
\Omega^{\hat \imath}{}_{\hat \jmath} &=& - \sum_{k \ne i,j} (\partial_k
\gamma_{ji} - \gamma_{ki} \gamma_{jk}) \, dz^i \wedge dz^k 
- \sum_{k \ne i,j} (\partial_k \gamma_{ij} - \gamma_{ik} \gamma_{kj})
\, dz^k \wedge dz^j \nonumber \\
&& \hskip 6.em{} - (\partial_i \gamma_{ij} + \partial_j \gamma_{ji} +
\sum_{k \ne i,j} \gamma_{ki} \gamma_{kj}) \, dz^i \wedge dz^j.
\label{1.9} 
\end{eqnarray}

Using (\ref{1.9}) we conclude that the Riemannian submanifold $(U,
g|_U)$ of the Riemannian manifold $(M, g)$ is flat if and only if the
rotation coefficients $\gamma_{ij}$ satisfy the following system of
partial differential equations
\begin{eqnarray}
&&\partial_k \gamma_{ij} = \gamma_{ik} \gamma_{kj}, \quad i \ne j \ne
k, \label{1.1} \\
&&\partial_i \gamma_{ij} + \partial_j \gamma_{ji} + \sum_{k \neq i,j}
\gamma_{ki} \gamma_{kj} = 0, \quad i \neq j, \label{1.2}
\end{eqnarray}
where the notation $i \ne j \ne k$ means that $i$, $j$, $k$ are
distinct. 
{}From the other hand, let $(U; z^1, \ldots, z^n)$ be a chart on the
manifold $M$, and we have a solution $\gamma_{ij}$ of equations
(\ref{1.1}), (\ref{1.2}). Let us rewrite (\ref{1.3}) in the form 
\[
\partial_i \beta_j = \gamma_{ij} \beta_i,
\]
and consider these equalities as equations for the functions
$\beta_i$. It is easy to show that due to (\ref{1.1}), the
integrability conditions for these equations are satisfied, and we
can find the functions $\beta_i$ which play the role of the Lam\'e
coefficients having $\gamma_{ij}$ as the corresponding rotation
coefficients. If we define a metric tensor on $U$ by
\begin{equation}
g = \sum_{i=1}^n \beta_i^2 \, dz^i \otimes dz^i, \label{1.10}
\end{equation}
then submanifold $U$ becomes a flat Riemannian manifold whose metric
is diagonal with respect to the coordinates $z^i$.

Equations (\ref{1.1}), (\ref{1.2}) are called the {\it Lam\'e
equations}.  With the so called {\it Egoroff property},
$\gamma_{ij}=\gamma_{ji}$, equations (\ref{1.2}) are equivalent to
the following ones:
\begin{equation}
\left( \sum_{k=1}^n \partial/\partial_k \right) \gamma_{ij} = 0,
\qquad i \ne j.\label{EC}
\end{equation}
The corresponding solutions are represented in the form
$\beta_i^2=\partial_i F$ where $F$ is some function of the
coordinates $z^i$. The system consisting of equations (\ref{1.1}) and
(\ref{EC}) is called sometimes the {\it Darboux-Egoroff equations}.

If $(U, g|_U)$ is a Riemannian manifold of constant
curvature with the sectional curvature $k$, then we have \cite{KNo63}
\[
\Omega^{\hat \imath}{}_{\hat \jmath} = k \, \theta^{\hat \imath} \wedge
\theta^{\hat \jmath} = k \, \beta_i \beta_j \, dz^i \wedge dz^j.
\]
Therefore, taking into account (\ref{1.9}), one sees that the
Riemannian submanifold $(U, g|_U)$ is of constant curvature with the
sectional curvature $k$ if and only if the rotation coefficients
satisfy the equations
\begin{eqnarray}
&&\partial_k \gamma_{ij} = \gamma_{ik} \gamma_{kj}, \quad i \ne j \ne
k, \label{1.5} \\
&&\partial_i \gamma_{ij} + \partial_j \gamma_{ji} + \sum_{k \neq i,j}
\gamma_{ki} \gamma_{kj} + k \beta_i \beta_j = 0, \quad i \neq j.
\label{1.6} 
\end{eqnarray}
We call equations (\ref{1.5}), (\ref{1.6}) and (\ref{1.3}) the {\it
Bourlet type equations}. The Bourlet equations in the precise sense
correspond to the case with $k=1$ and $\sum_{i=1}^n \beta_i^2 = 1$,
see, for example, \cite{Bia24,Dar10}.

Sometimes it is suitable to rewrite at least a part of equations
(\ref{1.5}), (\ref{1.6}) in a `Laplacian' type form. Impose the
condition
\begin{equation}
k \sum_{i=1}^n \beta_i^2 = c,
\label{1.7}
\end{equation}
where $c$ is a constant. It is convenient to allow the functions
$\beta_i$, and hence the functions $\gamma_{ij}$, to take complex
values. Therefore, we will assume that $c$ is an arbitrary complex
number.  One can easily get convinced by a direct check with account
of (\ref{1.3}) and (\ref{1.7}) that there takes place the relation
\[
\partial_i\beta_i = - \sum_{j \neq i} \gamma_{ij} \beta_j.
\]
Now, using the same calculations as those in \cite{Ami81,Sav86}, and 
introducing, as there, the operators
\[
\Delta_{(i)} = \sum_{j \ne i} \partial_j^2 -\partial_i^2, 
\]
we obtain from equations (\ref{1.6}) 
\begin{eqnarray*}
\Delta_{(i)} \beta_i = \sum_{j \ne i} \beta_i [(\beta_i^{-1}
\partial_i \beta_j)^2 &-& (\beta_j^{-1} \partial_j \beta_i)^2] \\
&-& 2\sum_{j \ne k \ne i} \beta_j \beta_k^{-2} (\partial_k \beta_i)
(\partial_k \beta_j) + \beta_i (k \beta_i^2 - c).
\end{eqnarray*}

For $n = 2$ equations (\ref{1.5}) and (\ref{1.1}) are absent;
equation (\ref{1.6}) is reduced to the Liouville and sine-Gordon
equations for $\beta_1^2 + \beta_2^2$ equals $0$ and $1$,
respectively; while (\ref{1.2}) is the wave equation. This is why for
higher dimensions the Bourlet type equations with a nonzero constant
$c$ in (\ref{1.7}), with $c = 0$, and the Lam\'e equations are called
sometimes multidimensional generalisations of the sine-Gordon,
Liouville, and wave equations, respectively, see, for example,
\cite{Ami81,TTe80,Sav86,ABT86}.

In the beginning of eighties an interest to the Lam\'e and Bourlet
type equations was revived. In particular, it was shown that system
(\ref{1.5}), (\ref{1.6}), (\ref{1.7}) with $k=1$ and $c=1$ provides
the necessary and sufficient conditions for the construction of a
local immersion of the Lobachevsky space $L_n$ into ${\Bbb R}^{2n-1}$
\cite{Ami81}, see also \cite{TTe80}.  Further, as it follows from the
results of works \cite{Tsa91,Tsa93}, the problems of description of
$n$-orthogonal curvilinear coordinate systems and of the
classification of integrable Hamiltonian systems of hydrodynamic type
\cite{DNo83} are almost equivalent. Note that system (\ref{1.5}) is a
natural generalisation \cite{Zak96} of the three wave system which is
a relevant object in nonlinear optics.  The Lam\'e equations also
arise very naturally in the context of the Cecotti-Vafa equations
describing topological-antitopological fusion, see \cite{Dub93} and
references therein, and in those of the multidimensional
generalisations of the Toda type systems \cite{RSa96}.

Probably the most interesting modern area where the Lam\'e equations
with the corresponding initial conditions appear to be quite
relevant, is related to the theory of Frobenius manifolds in the
spirit of B.~A.~Dubrovin \cite{Dub93,Dub96}, N.~Hitchin \cite{Hit96}
and Yu.~I.~Manin \cite{Man96}. In particular, in the last very
remarkable paper semisimple Frobenius manifolds are related to
solutions of the Schlesinger equations, constrained by some special
initial conditions. In our notations it corresponds to solutions of
the Lam\'e system satisfying the Egoroff property, condition
(\ref{1.7}) with $c = 0$, which together automatically provide the
validity of (\ref{EC}), since here $(\sum_k \partial_k)
\beta_i = 0$; and also a rather restrictive requirement which is
equivalent to $(\sum_k z^k \partial_k) \beta_i \sim \beta_i$.

Finally note that the classification and description of diagonal
metrics seems to be relevant for some modern problems of supergravity
theories, including their elementary and solitonic supersymmetric
$p$-brane solutions, see, for example, \cite{LPSS95} and references
therein. 

The very fact of integrability of the equations in question has been
established for quite a long time ago; the general solution is
defined by $n(n-1)/2$ functions of two variables for the Lam\'e
system, and by $n(n-1)$ functions of one variable and $n$ constants
for the Bourlet system.  However, an explicit form of the solutions
for higher dimensions remained unknown.

In the present work we obtain in an explicit and rather simple form
some special class of the solutions to the Lam\'e equations and to
the Bourlet type equations with and without condition (\ref{1.7}).
If one does not impose condition (\ref{1.7}), then our solutions are
determined by $n$ arbitrary functions of one variable, while
with condition (\ref{1.7}) the obtained solutions of the Bourlet
equations are expressed as products of Jacobi elliptic functions and
are determined by $2n$ arbitrary constants. The derivation of the
solutions to both of these systems is given by using two different
methods. One is based on the geometrical interpretation of the
corresponding equations.  Another approach uses a zero curvature (Lax
type) representation of the Lam\'e and Bourlet type equations.

The Lax type representations of the Lam\'e and Bourlet type
equations, different from ours, were considered in \cite{Zak96} and
\cite{ABT86}, respectively; see also \cite{Dub90}. In particular, the
author of \cite{Zak96}, using a multidimensional generalisation of
the Zakharov-Shabat dressing method \cite{ZSh74}, succeeded to obtain
some explicit solutions of the Lam\'e equations parametrised by $n$
functions of one variable, which are in a sense complementary to
those presented below.  In the beginning of November 1996, we were
informed by  V.~E.~Zakharov that he extended the results of
\cite{Zak96} to the Bourlet system.

\section{Bourlet type equations}

We begin with the description of the zero curvature representation of
the Bourlet type equations following \cite{Sav86}.  Consider 
the case $k > 0$. Here the zero curvature representation
is based on the Lie group ${\rm O}(n+1, {\Bbb R})$.  For the case $k
< 0$ we should use the Lie group ${\rm O}(n, 1)$. Actually, we can
complexify the Bourlet equations allowing the functions $\beta_i$ and
$\gamma_{ij}$ to take complex values. In such a case we should use
for the construction of the zero curvature representation the complex
Lie group ${\rm O}(n+1, {\Bbb C})$ and here positive and negative $k$
can be considered simultaneously. It is clear that without any loose
of generality we can take $k=1$.

Let ${\sf M}_{ab}$ be the elements of the Lie algebra ${\frak o}(n+1,
{\Bbb R})$ of the Lie group ${\rm O}(n+1, {\Bbb R})$ defined as
\[
({\sf M}_{ab})_{cd} = \delta_{ac} \delta_{bd} - \delta_{bc} \delta_{ad},
\]
The commutation relations for these elements have the standard form
\[
[{\sf M}_{ab}, {\sf M}_{cd}] = \delta_{ad} {\sf M}_{bc} + \delta_{bc}
{\sf M}_{ad} - \delta_{ac} {\sf M}_{bd} - \delta_{bd} {\sf M}_{ac},
\]
and any element ${\sf X}$ of ${\frak g}$ can be represented as
\[
{\sf X} = \sum_{a,b =1}^{n+1} x_{ab} {\sf M}_{ab}.
\]
Such a representation is unique if we suppose that $x_{ab} = -
x_{ba}$.

In what follows we assume that the indices $a, b, c, \ldots$ run from
$1$ to $n+1$, while the indices $i, j, k, \ldots $ run from $1$ to
$n$. Let $(U; z^1, \ldots, z^n)$ be a chart on some smooth manifold
$M$.  Consider the connection $\omega = \sum_{i = 1}^n \omega_i dz^i$
on the trivial principal fibre bundle $U \times {\rm O}(n+1, {\Bbb
R})$ with the components given by
\begin{equation}
\omega_i = \sum_{k=1}^n \gamma_{ki} {\sf M}_{ik} + \beta_i {\sf
M}_{i, n+1}. \label{2.2} 
\end{equation}
One can get convinced that the Bourlet type equations (\ref{1.3}),
(\ref{1.5}) and (\ref{1.6}) are equivalent to the zero curvature
condition for the connection $\omega$, which, in terms of the
connection components, has the form
\begin{equation}
\partial_i \omega_j- \partial_j \omega_i + [\omega_i, \omega_j] =
0. \label{2.3}
\end{equation}

Identify the Lie group ${\rm O}(n, {\Bbb R})$ with the Lie subgroup
of ${\rm O}(n+1, {\Bbb R})$ formed by the matrices ${\sf A} \in {\rm
O}(n+1, {\Bbb R})$, such that
\[
{\sf A}_{i, n+1} = 0, \qquad {\sf A}_{n+1, j} = 0, \qquad {\sf
A}_{n+1, n+1} = 1. 
\]
Similarly, identify the Lie algebra ${\frak o}(n, {\Bbb R})$ with the
corresponding subalgebra of ${\frak o}(n+1, {\Bbb R})$. 

Let the connection $\omega$ with the components of form (\ref{2.2})
satisfies the zero curvature condition (\ref{2.3}). Suppose that $U$ is
simply connected, then there exists a mapping $\varphi$ from $U$ to
${\rm O}(n+1, {\Bbb R})$, such that
\[
\omega_i = \varphi^{-1} \partial_i \varphi.
\]
Parametrise $\varphi$ in the following way
\begin{equation}
\varphi = \xi \chi, \label{2.10}
\end{equation}
where $\chi$ is a mapping from $U$ to ${\rm O}(n, {\Bbb R})$ and the
mapping $\xi$ has the form 
\begin{equation}
\xi = e^{\psi_1 {\sf M}_{12}} e^{\psi_2 {\sf M}_{23}} \cdots
e^{\psi_{n-1} {\sf M}_{n-1,n}} e^{\psi_n {\sf M}_{n,n+1}}.
\label{2.4}
\end{equation}
Here $\psi_i$ are some functions on $U$ having the meaning of the
generalised Euler angles \cite{Vil69}. For the connection components
$\omega_i$ one obtains the expression
\[
\omega_i = \chi^{-1} (\xi^{-1} \partial_i \xi) \chi + \chi^{-1}
\partial_i \chi.
\]
Relation (\ref{2.4}) gives
\[
\xi^{-1} \partial_i \xi = \sum_{j=1}^{n-1} \partial_i \psi_j \sum_{k
= j+1}^n \mu_{jk} (\psi) \, {\sf M}_{jk} + \sum_{j=1}^n \partial_i \psi_j
\, \nu_j (\psi) \, {\sf M}_{j, n+1},
\]
where
\begin{eqnarray}
&&\mu_{j-1, j}(\psi) = \cos \psi_j, \quad 1 < j \le n, \label{2.30} \\ 
&&\mu_{jk}(\psi) = \left( \prod_{l=j+1}^{k-1} \sin \psi_l \right)
\cos \psi_k, \quad 1 < j+1 < k \le n, \label{2.31} \\
&&\nu_j (\psi) = \prod_{l =j+1}^n \sin \psi_l, \quad 1  \le j < n, \qquad
\nu_n(\psi) = 1. \label{2.32}
\end{eqnarray}
Now, using the evident equalities
\begin{equation}
\chi^{-1} \partial_i \chi = \frac{1}{2} \sum_{j,k,l =1}^n \chi_{lj}
\partial_i \chi_{lk} {\sf M}_{jk}, \qquad \chi^{-1} {\sf M}_{i, n+1}
\chi = \sum_{j=1}^n \chi_{ij} {\sf M}_{j, n+1},\label{2.14} 
\end{equation}
one comes to the expressions
\begin{eqnarray}
\omega_i &=& \frac{1}{2} \sum_{j,k,l=1}^n \chi_{lj} \, \partial_i
\chi_{lk} \, {\sf M}_{jk} \nonumber \\
&+& \sum_{j,k=1}^n \sum_{l=1}^{n-1} \partial_i \psi_l
\sum_{m = l+1}^n \mu_{lm}(\psi) \, \chi_{lj} \, \chi_{mk} \, {\sf
M}_{jk} + \sum_{j,l=1}^n \partial_i \psi_l \, \nu_l(\psi) \,
\chi_{lj} \, {\sf M}_{j, n+1}. \hskip 2em \label{2.5} 
\end{eqnarray}
Comparing (\ref{2.5}) and (\ref{2.2}), we have, in particular,
\begin{equation}
\sum_{l=1}^n \partial_i \psi_l \, \nu_l(\psi) \, \chi_{lj} =
\beta_i \, \delta_{ij}.
\label{2.6} 
\end{equation}
Note that the geometrical meaning of the functions $\beta_i$ do not
allow them to take zero value. Therefore, from (\ref{2.6}) it follows
that for any point $p \in U$ we have
\begin{equation}
\det (\partial_i \psi_j(p)) \ne 0, \qquad  \nu_i(\psi(p)) \ne 0.
\label{2.8} 
\end{equation}
Since the matrix $(\chi_{ij})$ is orthogonal, one easily obtains
\[
\chi_{ij} = \frac{1}{\beta_j} \partial_j \psi_i \, \nu_i(\psi),
\]
and, using again the orthogonality of $(\chi_{ij})$, one sees that 
\begin{equation}
\beta_i^2 = \sum_{l=1}^n (\partial_i \psi_l \, \nu_l(\psi))^2.
\label{2.29} 
\end{equation}
Therefore, we have
\begin{equation}
\chi_{ij} = \frac{\partial_j \psi_i \, \nu_i(\psi)}{\sqrt{\sum_{l=1}^n
(\partial_j \psi_l \, \nu_l(\psi))^2}}. \label{2.9}  
\end{equation}
Thus, the matrix $(\chi_{ij})$, and hence the mapping $\chi$, is
completely determined by the functions $\psi_i$, and its
orthogonality is equivalently realised by the relation
\begin{equation}
\sum_{l=1}^n \partial_i \psi_l \, \nu_l^2(\psi) \, \partial_j \psi_l
= 0, \qquad i \ne j. 
\label{2.7}
\end{equation}

Suppose now that a set of functions $\psi_i$ satisfies relations
(\ref{2.8}) and (\ref{2.7}). Consider the mapping $\varphi$ defined
by (\ref{2.10}) with the mapping $\xi$ having form (\ref{2.4}) and
the mapping $\chi$ defined by (\ref{2.9}). Show that the mapping
$\varphi$ generates the connection with the components of form
(\ref{2.2}).  First of all, with $\beta_i$ of form (\ref{2.29}) we
can get convinced that in the case under consideration relation
(\ref{2.6}) is valid. Taking into account (\ref{2.7}), one can write
the relation
\[
\sum_{l=1}^n \partial_j \psi_l \, \nu_l^2 (\psi) \, \partial_k \psi_l
= \beta^2_j \, \delta_{jk}, 
\] 
whose differentiation with respect to $z^i$ gives
\begin{eqnarray*}
&&\sum_{l=1}^n \partial_j \psi_l \, \nu_l^2(\psi) \, \partial_i
\partial_k \psi_l \\
&&\hskip 2em {} = - \sum_{l=1}^n \partial_i \partial_j \psi_l \,
\nu_l^2(\psi) \, \partial_k \psi_l - 2 \sum_{l=1}^n \partial_j
\psi_l \, \nu_l(\psi) \, \partial_i \nu_l(\psi) \, \partial_k \psi_l
+ 2 \beta_j \partial_i \beta_j \, \delta_{jk}.
\end{eqnarray*}
Since the left hand side of this equality is symmetric with respect
to the transposition of the indices $i$ and $k$, its right hand side
must also be symmetric with respect to this transposition, and,
therefore, we have 
\begin{eqnarray*}
&&\sum_{l=1}^n \partial_j \psi_l \, \nu_l^2(\psi) \, \partial_i
\partial_k \psi_l \\
&& \hskip 2em {} = - \sum_{l=1}^n \partial_k \partial_j \psi_l \,
\nu_l^2(\psi) \, \partial_i \psi_l - 2 \sum_{l=1}^n \partial_j
\psi_l \, \nu_l(\psi) \, \partial_k \nu_l(\psi) \, \partial_i \psi_l
+ 2 \beta_j 
\partial_k \beta_j \, \delta_{ij}.
\end{eqnarray*}
Using this equality, it is not difficult to show that
\begin{eqnarray}
&&\sum_{l=1}^n \chi_{lj} \, \partial_i \chi_{lk} = \gamma_{kj} \,
\delta_{ij} - \gamma_{jk} \, \delta_{ik} \nonumber \\*
&&\hskip 1.em {} - \frac{1}{\beta_j \beta_k} \sum_{l=1}^n [\partial_j
\psi_l \, \nu_l(\psi) \, \partial_k \nu_l(\psi) \, \partial_i \psi_l -
\partial_k \psi_l \, \nu_l(\psi) \,
\partial_j \nu_l(\psi) \, \partial_i \psi_l], \hskip 2.em \label{2.11} 
\end{eqnarray}
where the functions $\gamma_{ij}$ are defined by (\ref{1.3}).

Using the concrete form of the functions $\mu_{ij}(\psi)$ and
$\nu_i(\psi)$, we can get convinced in the validity of the equalities
\[
\frac{\partial \nu_j(\psi)}{\partial \psi_i} = 0, \quad 1 \le i
\le j, \qquad \mu_{ij}(\psi) = \frac{1}{\nu_j(\psi)} \frac{\partial
\nu_i(\psi)}{\partial \psi_j},\]
which allow to show that 
\begin{equation}
\sum_{l=1}^{n-1} \partial_i \psi_l \sum_{m = l+1}^n
\mu_{lm}(\psi) \, \chi_{lj} \, \chi_{mk} = \frac{1}{\beta_j \beta_k}
\sum_{l=1}^n \partial_j \psi_l \, \nu_l(\psi) \, \partial_k \nu_l(\psi) \,
\partial_i \psi_l. \label{2.23}
\end{equation}

Substituting (\ref{2.11}), (\ref{2.23}) and (\ref{2.6}) into
(\ref{2.5}), we come to expression (\ref{2.2}). Thus, any set of
functions $\psi_i$ satisfying (\ref{2.8}) and (\ref{2.7}) allows to
construct a connection of form (\ref{2.2}) satisfying the zero
curvature condition (\ref{2.3}) which is equivalent to the Bourlet
type equations. Therefore, the general solution to the
Bourlet type equations is described by (\ref{2.29}) where the
functions $\psi_i$ satisfy (\ref{2.8}) and (\ref{2.7}). In the
simplest case we can satisfy (\ref{2.7}) assuming that
\begin{equation}
\partial_i \psi_j = 0, \qquad i \ne j; \label{2.33}
\end{equation}
in other words, each function $\psi_i$ depends on the corresponding
coordinate $z^i$ only. In this case we obtain the following
expressions for the functions $\beta_i$:
\begin{equation}
\beta_i = \partial_i \psi_i \prod_{j=i+1}^n \sin \psi_j, \quad
1 \leq i < n, \qquad \beta_n = \partial_n \psi_n. \label{2.13}
\end{equation}
The corresponding expressions for the functions $\gamma_{ij}$ can be
easily found and we do not give here their explicit form. Note here
that since $\gamma_{ij} = 0$ for $i < j$, our solutions do not
satisfy the Egoroff property.

There is a transparent geometrical interpretation of the results
obtained above.  Recall that solutions of the Bourlet type equations
are associated with diagonal metrics in Riemannian spaces of constant
curvature.  Namely, let $(M, g)$ be a Riemannian manifold of constant
curvature with the sectional curvature $k$, and $(U; z^1, \ldots,
z^n)$ be such a chart on $M$ that the metric $g$ has on $U$ form
(\ref{1.4}). Then the Lam\'e and the corresponding rotation
coefficients satisfy the Bourlet type equations. From the other
hand, let $(U; z^1, \ldots, z^n)$ be a chart on a manifold $M$, and
we have a solution of the Bourlet type equations.  Supply the open
submanifold $U$ with metric (\ref{1.10}); then $(U, g)$ becomes a
Riemannian manifold of constant curvature with the sectional
curvature $k$.

The simplest example of a manifold of constant curvature is an 
$n$-dimen\-si\-onal sphere in ${\Bbb R}^{n+1}$ with the metric
induced by the standard metric on ${\Bbb R}^{n+1}$. Here if the
radius of the sphere is $R$, then the sectional curvature is $1/R^2$.
Let us show that the corresponding metric is diagonal with respect to
the spherical coordinates.  

Begin with the consideration of the standard metric in ${\Bbb R}^n$.
Denoting the standard coordinates on ${\Bbb R}^n$ by $x_i$ and the
spherical coordinates by $r$ and $\theta_1, \ldots \theta_{n-1}$, one
has
\[
x^1 = r \prod_{k=1}^{n-1} \sin \vartheta_k,\quad x^i = r \cos
\vartheta_{i-1} \prod_{k = i}^{n-1} \sin \vartheta_k,\; 1<i<n,\quad
x^n = r \cos \vartheta_{n-1}. 
\]
The standard metric on ${\Bbb R}^n$ has the form
\[
\g{n} = \sum_{l=1}^n dx_l \otimes dx_l.
\]
Denote the functions describing the dependence of the coordinates
$x^i$ on $r$ and $\theta_1,\ldots,\theta_{n-1}$, by $\f{i}{n}(r,
\theta)$. For any $n$ 
one has
\begin{equation}
\sum_{l=1}^n \f{l}{n}\negthinspace{}^2 = r^2. \label{a.1}
\end{equation}
Taking external derivative of this equality, we obtain
\begin{equation}
\sum_{l=1}^n \f{l}{n} \, d\f{l}{n} = r \, dr. \label{a.2}
\end{equation}
It is easy to get convinced that 
\[
\f{i}{n} = \f{i}{n-1} \sin \theta_{n-1}, \quad 1 \le i < n, \qquad
\f{n}{n} = r \cos \theta_{n-1}. 
\]
These relations imply
\begin{eqnarray*}
&&d\f{i}{n} = \sin \theta_{n-1} \, d\f{i}{n-1} + \f{i}{n-1} \cos
\theta_{n-1} \, d \theta_{n-1}, \quad 1 \le i < n, \\
&&d\f{n}{n} = \cos \theta_{n-1} \, dr - r \sin \theta_{n-1} \, d
\theta_{n-1}. 
\end{eqnarray*}
Substituting these equalities into the relation
\[
\g{n} = \sum_{l=1}^n d\f{l}{n} \otimes d\f{l}{n}
\]
and using (\ref{a.1}), (\ref{a.2}) we obtain
\[
\g{n} = \left( \g{n-1} - dr \otimes dr \right) \sin^2 \theta_{n-1} +
r^2 d\theta_{n-1} \otimes d\theta_{n-1} + dr \otimes dr.
\]
This equality gives
\begin{equation}
\g{n} = r^2 \left[ \sum_{l=1}^{n-2} \left( \prod_{m=l+1}^{n-1}
\sin^2 \theta_m \right) d\theta_l \otimes d\theta_l + d\theta_{n-1}
\otimes d\theta_{n-1} \right] + dr \otimes dr. \label{2.36}
\end{equation}

Consider now the unit $n$-dimensional sphere $S^n$ in ${\Bbb R}^{n+1}$.
Denote the spherical coordinates in $S^n$ by $z^1, \ldots, z^n$. As
it follows from (\ref{2.36}), the explicit expression for the metric on
$S^n$ in terms of the spherical coordinates has the form
\[
g = \sum_{l=1}^{n-1} \left( \prod_{m=l+1}^n \sin^2 z^m \right)
dz^l \otimes dz^l + dz^n \otimes dz^n.
\]
So we have a diagonal metric. Note that it can be written in the form
\begin{equation}
g = \sum_{l=1}^n \nu^2_l(z) \, dz^l \otimes dz^l, \label{2.34}
\end{equation}
where the functions $\nu_i$ are given by (\ref{2.32}). Let $\psi$ be
a diffeomorphism from $S^n$ to $S^n$. It is clear that $(S^n, \psi^*
g)$ is also a Riemannian manifold of constant curvature with the
sectional curvature equal to $1$. Denoting $\psi^* z^i = \psi_i$, one
gets
\[
\psi^* g = \sum_{j,k,l=1}^n \partial_j \psi_l \, \nu^2_l(\psi) \,
\partial_k \psi_l \, dz^j \otimes dz^k.
\]
Therefore, the metric $\psi^* g$ is diagonal with respect to the
coordinates $z^i$ if and only if the functions $\psi_i$ satisfy
relations (\ref{2.7}). In particular, if the functions $\psi_i$
satisfy relations (\ref{2.33}) we obtain the diagonal metrics with the
Lam\'e coefficients given by (\ref{2.13}).

In general, starting from some fixed diagonal metric in the space of
constant curvature with the unit sectional curvature, one gets the
family of explicit solutions to the Bourlet type equations
parametrised by a set of $n$ functions each depending only on one
variable.  In terms of equations (\ref{1.3}), (\ref{1.5}) and
(\ref{1.6}) themselves, we formulate this observation as follows. Let
the functions $\beta_i$, $\gamma_{ij}$ satisfy the Bourlet type
equations; then for any set of functions $\psi_i$, such that
\[
\partial_i \psi_j = 0, \qquad i \ne j,
\]
the functions
\begin{equation}
\beta'_i(z) = \beta_i(\psi(z)) \, \partial_i \psi_i(z), \qquad
\gamma'_{ij} (z) = \gamma_{ij} (\psi(z)) \, \partial_j \psi_j(z^j),
\label{2.35} 
\end{equation}
where $\psi(z)$ stands for the set $\psi_1(z), \ldots \psi_n(z)$,
also satisfy the Bourlet type equations. 

Note that our considerations can be easily generalised to the case of
complex metrics. In this case the zero curvature representation of
the Bourlet type equations should be based on the Lie group ${\rm
O}(n+1, {\Bbb C})$.

Return to the consideration of solutions (\ref{2.13}) to the Bourlet 
type equations.  If one imposes condition
(\ref{1.7}) where $c$ is an arbitrary zero or nonzero constant, then
the arbitrary functions $\psi_i (z^i)$ satisfy the equation
\[
\sum_{l = 1}^n \left( \prod_{m=l+1}^n \sin^2 \psi_m \right) (\partial_l \psi_l)^2
= c,
\]
thereof for some constants $c_i$, $i = 0, \ldots, n$, such that $c_0
= 0$ and $c_n = c$, one gets
\begin{equation}
\partial_i \psi_i= \sqrt{c_i - c_{i-1} \sin^2 \psi_i}. \label{el1}
\end{equation}
Hence, solution (\ref{2.13}) takes the form
\begin{equation}
\beta_i =  \sqrt{c_i - c_{i-1} \sin^2 \psi_i} \prod_{j=i+1}^n \sin
\psi_j, \label{a44}
\end{equation}
where the functions $\psi_i$ are determined by the ordinary
differential equations (\ref{el1}). Suppose that all constants $c_i$,
$i = 1,\ldots,n$, are different  from zero. With appropriate
conditions on the constants $c_i$, in accordance with (\ref{el1}) one
has
\[
z^i + d^i = \int^{\psi_i}_0 \frac{d\psi_i}{\sqrt{c_i-c_{i-1} \sin^2
\psi_i}}, 
\]
where $d^i$ are arbitrary constants. Therefore,
\[
\sqrt{c_i}(z^i + d^i) = F \left( \psi_i, \,
\sqrt{\frac{c_{i-1}}{c_i}}\, \right),
\]
where $F(\phi, k)$ is the elliptic integral of the first kind,
\[
F(\phi, k)=\int^{\phi}_0 \frac{d\phi}{\sqrt{1 - k^2 \sin^2 \phi}}.
\]
Thus, using Jacobi elliptic functions, we can write
\begin{eqnarray*}
&&\sin \psi_i(z^i) = \sn \left( \sqrt{c_i} (z^i+d^i), \,
\sqrt{\frac{c_{i-1}}{c_i}} \, \right), \\
&&\cos \psi_i(z^i) = \cn \left( \sqrt{c_i} (z^i+d^i), \,
\sqrt{\frac{c_{i-1}}{c_i}} \, \right). 
\end{eqnarray*}
Now, with the evident relation
\[
\partial_i \psi_i(z^i) = \frac{\partial_i \sin \psi_i(z^i)}{\cos
\psi_i(z^i)} = \sqrt{c_i} \dn \left( \sqrt{c_i} (z^i+d^i), \,
\sqrt{\frac{c_{i-1}}{c_i}} \, \right),
\]
we write our solution as the product of elliptic functions,
\begin{equation}
\beta_i(z) = \sqrt{c_i} \dn \left( \sqrt{c_i} (z^i+d^i), \,
\sqrt{\frac{c_{i-1}}{c_i}} \, \right) \prod_{j=i+1}^n \sn \left( \sqrt{c_j}
(z^j+d^j), \, \sqrt{\frac{c_{j-1}}{c_j}} \, \right).\label{rez}
\end{equation}
The case when some of the constants $c_i$ are equal to zero can be
analysed in a similar way.
Note that, taking into account the relations
\[
\sn (u, 1) = \tanh u, \quad \dn(u, 1) = \frac{1}{\cosh u}, \quad \sn
(u, 0) = \sin u, \quad \dn (u, 0) = 1, 
\]
with an appropriate choice of the constants $c_i$, we can reduce some
of the elliptic functions entering the obtained solution to the
trigonometric or hyperbolic ones.

It is clear from the solution in form (\ref{a44}) or (\ref{rez}),
that it does not depend on the variable $z^1$ at all, since among
the functions $\beta_i$, only $\beta_1$ depends on $\psi_1$ and only
as $\partial_1\psi_1$, while $\psi_1= c_1 z^1+d^1$.

In the simplest case $n=2$ and $c_2=1$ with the parametrisation
$\beta_1 = \cos(u/2)$, $\beta_2 = \sin(u/2)$, system (\ref{1.3}),
(\ref{1.5}) and (\ref{1.6}) is reduced to the sine-Gordon equation
\[
\partial_1^2 u - \partial_2^2 u + \sin u = 0,
\]
and one gets the evident solution $\sin(u/2) = \dn (z^2 + d^2, \sqrt{c_1})$.

\section{Lam\'e equations}

The zero curvature representation of the Lam\'e equations is based on
the Lie group $G$ of rigid motions of the affine space ${\Bbb R}^n$.
This Lie group is isomorphic to the semidirect product of the Lie
groups ${\rm O}(n, {\Bbb R})$ and ${\Bbb R}^n$, where the linear
space ${\Bbb R}^n$ is considered as a Lie group with respect to the
addition operation.  The standard basis of the Lie algebra ${\frak
g}$ of the Lie group $G$ consists of the elements ${\sf M}_{ij}$ and
${\sf P}_i$ which satisfy the commutation relations
\begin{eqnarray*}
&[{\sf M}_{ij}, {\sf M}_{kl}] = \delta_{il} {\sf M}_{jk} +
\delta_{jk} {\sf M}_{il} - \delta_{ik} {\sf M}_{jl} - \delta_{jl}
{\sf M}_{ik},& \\ 
&[{\sf M}_{ij}, {\sf P}_k] = \delta_{jk} {\sf P}_i - \delta_{ik} {\sf
P}_j, \qquad [{\sf P}_i, {\sf P}_j] = 0.
\end{eqnarray*}

Let $(U; z^1, \ldots, z^n)$ be a chart on the manifold $M$.  Consider
the connection $\omega = \sum_{i = 1}^n \omega_i dz^i$ on the trivial
principal fibre bundle $U \times G$ with the components given by
\begin{equation}
\omega_i = \sum_{k=1}^n \gamma_{ki} {\sf M}_{ik} + \beta_i {\sf P}_i.
\label{4.2} 
\end{equation}
It can be easily verified that equations (\ref{1.3})--(\ref{1.2}) are
equivalent to the zero curvature condition for the connection
$\omega$. It is well known that the Lie algebra ${\frak g}$ can be
obtained from the Lie algebra ${\frak o}(n+1, {\Bbb R})$ by an
appropriate In\"on\"u-Wigner contraction. Unfortunately, this fact
does not give us a direct procedure for obtaining solutions of the
Lam\'e equations from solutions of the Bourlet type equations.
Therefore, we will consider the procedure for obtaining solutions of
the Lam\'e equations independently.

Let the connection $\omega$ with the components of form (\ref{4.2})
satisfies the zero curvature condition. Restricting to the case of
simply connected $U$, write for the connection components $\omega_i$
the representation
\[
\omega_i = \varphi^{-1} \partial_i \varphi,
\]
where $\varphi$ is some mapping from $U$ to $G$. Parametrise
$\varphi$ in the following way:
\begin{equation}
\varphi = \xi \chi \label{4.10}
\end{equation}
where $\chi$ is a mapping from $U$ to ${\rm O}(n, {\Bbb R})$,
and the mapping $\xi$ has the form
\begin{equation}
\xi = e^{\psi_1 {\sf P}_1} e^{\psi_2 {\sf P}_2} \cdots e^{\psi_{n-1}
{\sf P}_{n-1}} e^{\psi_n {\sf P}_n}. \label{4.4} 
\end{equation}
For the connection components $\omega_i$ one obtains the expression
\begin{equation}
\omega_i = \frac{1}{2} \sum_{j,k,l=1}^n \chi_{lj} \partial_i
\chi_{lk} {\sf M}_{jk} + \sum_{j,l=1}^n \partial_i \psi_l \chi_{lj}
{\sf P}_j. 
\label{4.5} 
\end{equation}
{}From the comparison of (\ref{4.5}) and (\ref{4.2}) we see that
\begin{equation}
\chi_{ij} = \frac{\partial_j \psi_i}{\sqrt{\sum_{l=1}^n (\partial_j
\psi_l)^2}}, \label{4.9}  
\end{equation}
and the functions $\psi_i$ satisfy the relation
\begin{equation}
\sum_{l=1}^n \partial_i \psi_l \, \partial_j \psi_l = 0, \qquad i \ne j.
\label{4.7}
\end{equation}
The functions $\beta_i$ are connected with the functions $\psi_i$ by
the formula
\begin{equation}
\beta_i^2 = \sum_{l=1}^n (\partial_i \psi_l)^2, \label{4.6}
\end{equation}
and from the geometrical meaning of $\beta_i$ it follows that
\begin{equation}
\det (\partial_i \psi_j(a)) \ne 0. \label{4.8}
\end{equation}

Suppose now that a set of functions $\psi_i$ satisfies relations
(\ref{4.7}) and (\ref{4.8}). Consider the mapping $\varphi$ defined
by (\ref{4.10}) with the mapping $\xi$ having form (\ref{4.4}) and
the mapping $\chi$ defined by (\ref{4.9}). It can be shown that the
mapping $\varphi$ generates the connection with the components of
form (\ref{4.2}). Here the functions $\beta_i$ are defined from
(\ref{4.6}), and the functions $\gamma_{ij}$ are given by
(\ref{1.3}). Thus, any set of functions $\psi_i$ satisfying
(\ref{4.8}) and (\ref{4.7}) allows to construct a connection of form
(\ref{4.2}) satisfying the zero curvature condition which is
equivalent to the Lam\'e equations, and in such a way we obtain the
general solution.

Assuming that the functions $\psi_i$ satisfy (\ref{2.33}), we have
\begin{equation}
\beta_i = \partial_i \psi_i. \label{4.11}
\end{equation}
It is clear that in this case $\gamma_{ij} = 0$. So one ends up with a
trivial solution of the Lam\'e equations. To get nontrivial solutions
one should consider different parametrisations of the mapping
$\varphi$.  For example, let us represent the mapping $\varphi$ in
form (\ref{4.10}) where the mapping $\chi$ again takes values in
${\rm O}(n, {\Bbb R})$, while the mapping $\xi$ has the form
\[
\xi = e^{\psi_1 {\sf M}_{12}} e^{\psi_2 {\sf M}_{23}} \cdots e^{\psi_{n-1}
{\sf M}_{n-1, n}} e^{\psi_n {\sf P}_n}. 
\]
With such a parametrisation of $\xi$, one gets
\begin{eqnarray*}
\omega_i &=& \frac{1}{2} \sum_{j,k,l=1}^n \chi_{lj} \, \partial_i
\chi_{lk} \, {\sf M}_{jk} \\
&+& \sum_{j,k=1}^n \sum_{l=1}^{n-1} \partial_i \psi_l
\sum_{m = l+1}^n \mu_{lm}(\psi) \, \chi_{lj} \, \chi_{mk} \, {\sf
M}_{jk} + \sum_{j,l=1}^n \partial_i \psi_l \, \nu_l(\psi) \,
\chi_{lj} \, {\sf P}_j, \hskip 2.em 
\end{eqnarray*}
where
\begin{eqnarray}
&&\mu_{j-1, j}(\psi) = \cos \psi_j, \quad 1 < j < n; \qquad
\mu_{n-1, n}(\psi) = 1; \label{4.30} \\
&&\mu_{jk}(\psi) = \left( \prod_{l=j+1}^{k-1} \sin \psi_l \right)
\cos \psi_k, \quad 1 < j+1 < k < n; \label{4.31} \\
&&\mu_{jn}(\psi) = \prod_{l=j+1}^{n-1} \sin \psi_l, \quad 1 < j+1 <
n; \label{4.32} \\ 
&&\displaystyle \nu_j (\psi) = \left( \prod_{k =j+1}^{n-1} \sin
\psi_k \right) \psi_n, \quad 1 \le j < n-1; \label{4.33a} \\
&&\nu_{n-1}(\psi) = \psi_n; \qquad \nu_n(\psi) = 1. \label{4.33b}
\end{eqnarray}
Using these relations we come to the following description of the
general solution to the Lam\'e equations. Let functions $\psi_i$
satisfy the relations
\[
\sum_{l=1}^n \partial_i \psi_l \, \nu_l^2(\psi) \, \partial_j \psi_l
= 0, \qquad i \ne j,
\]
and for any point $p \in U$ one has
\[
\det (\partial_i \psi_j(p)) \ne 0, \qquad  \nu_i(\psi(p)) \ne 0.
\]
Then the functions $\beta_i$ determined from the equality
\[
\beta_i^2 = \sum_{l=1}^n (\partial_i \psi_l \, \nu_l(\psi))^2,
\]
and the corresponding functions $\gamma_{ij}$ defined by (\ref{1.3})
give the general solution of the Lam\'e equations.
If the functions $\psi_i$ satisfy (\ref{2.33}), we get the following
expressions for the functions $\beta_i$
\begin{eqnarray}
&\beta_i = \partial_i \psi_i \left( \prod_{j=i+1}^{n-1}
\sin \psi_j \right) \psi_n, \quad 1 \leq i < n-1, \label{4.34a} \\
&\beta_{n-1} = \partial_{n-1} \psi_{n-1} \, \psi_n, \qquad \beta_n =
\partial_n \psi_n. \label{4.34b}
\end{eqnarray}

The geometrical interpretation of the obtained solutions to the
Lam\'e equations is similar to one given in the previous section.
Recall that solutions of the Lam\'e equations are associated with
flat diagonal metrics in flat Riemannian spaces. The simplest case
here is the standard metric in ${\Bbb R}^n$,
\[
g = \sum_{l=1}^n dz^l \otimes dz^l.
\]
Applying a diffeomorphism $\psi$ one gets the metric
\[
\psi^* g = \sum_{j,k,l=1}^n \partial_j \psi_l \partial_k \psi_l \, dz^j
\otimes dz^k,
\] 
which is diagonal if and only if the functions $\psi_i$ satisfy
(\ref{4.7}). Here the functions $\psi_i$ which obey (\ref{2.33}) give
the Lam\'e coefficients described by (\ref{4.11}).

A more nontrivial example is provided by the metric arising after the
transition to the spherical coordinates in ${\Bbb R}^n$. Introducing
the notations $z^i = \vartheta_i$, $i=1,\ldots,n-1$, and $z^n = r$,
we rewrite (\ref{2.36}) as
\[
g = (z^n)^2 \left[ \sum_{l=1}^{n-2} \left( \prod_{m=l+1}^{n-1}
\sin^2 z^m \right) dz^l \otimes dz^l + dz^{n-1} \otimes dz^{n-1}
\right] + dz^n \otimes dz^n. 
\]
Therefore, the metric $g$ has form (\ref{2.34}) with the functions
$\nu_i$ given by (\ref{4.33a}), (\ref{4.33b}). A diffeomorphism
$\psi$ with the functions $\psi_i = \psi^* z^i$ satisfying
(\ref{2.33}) gives the metric with the Lam\'e coefficients
(\ref{4.34a}), (\ref{4.34b}).

In conclusion note that relations (\ref{2.35}) describe the
symmetry transformations not only of the Bourlet type equations, but
also of the Lam\'e equations, and actually the existence of such
transformations allows us to construct the solutions of the equations
under consideration parametrised by $n$ arbitrary functions each
depending on one variable.

\section*{Acknowledgements}

The authors are grateful to B.~A.~Dubrovin, J.-L.~Gervais,
V.~A.~Kazakov, Yu.~I.~Manin and Yu.~G.~Stroganov for the discussions.
We also thank V.~E.~Zakharov for acquainting us with his recent
results prior to publication, and for some important comments. One of
the authors (M.~V.~S.) wishes to acknowledge the warm hospitality and
creative scientific atmosphere of the Laboratoire de Physique
Th\'eorique de l'\'Ecole Normale Sup\'erieure de Paris and of the
Max-Planck-Institut f\"ur Mathematik in Bonn. This work was
supported in part by the Russian Foundation for Basic Research under
grant no. 95--01--00125a.


\begin{thebibliography}{**}
\small

\bibitem{KNo63} 
Kobayashi, S., Nomizu K.: Foundations of differential geometry, Vol.
1. New York: Interscience, 1963.

\bibitem{Bia24}
Bianchi, L.: Lezioni di geometria differenziale, Vol. 2, Part 2. Bologna: 
Zanichelli 1924; Sisteme tripli ortogonali,  Opere, Vol. 3. Roma: Cremonese 
1955.

\bibitem{Dar10}
Darboux, G.: Le\c cons sur les syst\`emes orthogonaux et les coordonn\'ees
curvilignes. Paris: Gauthier-Villars 1910; Le\c cons sur la th\'eorie
g\'en\'erale des surfaces et les applications g\'eom\'etriques du calcul
infinit\'esimal, Vols. 1--4. Paris: Gauthier-Villars 1887--1896.

\bibitem{Ami81}
Aminov, Yu. A.: Isometric immersions of domains of $n$-dimensional 
Lo\-ba\-chev\-sky space in $(2n-1)$-dimen\-si\-onal euclidean space. Math.
USSR Sbornik {\bf 39(3)}, 359--386 (1981).

\bibitem{Sav86}
Saveliev, M. V.: Multidimensional nonlinear systems. Theor. Math. 
Phys. {\bf 69}, 1234--1240 (1986); Multidimensional nonlinear 
dynamical systems, in M. A.  Markov,  V. I.  Man'ko, V. V.  Dodonov
(eds.), ``Group Theoretical Methods in Physics'' Vol. 1, 113--127. Utrecht: 
VNU Science Press 1986.

\bibitem{TTe80} 
Tenenblat, K., Terng, C.-L.: B\"acklund theorem for $n$-dimensional
submanifolds of $R^{2n-1}$. Annals Math. {\bf 111}, 477--490 (1980); 
Terng, C.-L.: A higher dimensional generalization of the
Sine-Gordon equation and its soliton theory. Annals Math. {\bf 111}, 
491--510 (1980).

\bibitem{ABT86}
Ablowitz, M. J., Beals, R., Tenenblat, K.: On the solution of the generalized 
sine-Gordon equations. Stud. Appl. Math. {\bf 74}, 177--203 (1986).

\bibitem{Tsa91}
Tsarev, S. P.: The geometry of Hamiltonian systems of hydrodynamic
type.  The generalized hodograph method. Math. USSR Izvestija {\bf
37(2)}, 397-419 (1991).

\bibitem{Tsa93}
Tsarev, S. P.: Classical differential geometry and integrability of
systems of hydrodynamic type. Proc. NATO Advanced Research Workshop
on Applications of Analytical and Geometric Methods to Nonlinear
Differential Equations (Exeter, England, 14--19 July 1992),
hep-th/9303092 

\bibitem{DNo83}
Dubrovin, B. A., Novikov, S. P.:  Hamiltonian formalism of
one-dimensional systems of hydrodynamic type, and the
Bogolyubov-Whitham averaging method. Sov. Math. Doklady {\bf 27(3)},
665--669 (1983).

\bibitem{Zak96}
Zakharov, V. E.: Description of the $n$-orthogonal curvilinear coordinate
systems and Hamiltonian integrable systems of hydrodynamic type. Part
I. Integration of the Lam\'e equations. Submitted to Duke Math. Journal.

\bibitem{Dub93}
Dubrovin, B.: Geometry and integrability of topological-antitopological
fusion. Commun. Math. Phys. {\bf 152}, 539--564 (1993).

\bibitem{RSa96}
Razumov, A. V., Saveliev, M. V.: Multidimensional Toda type systems. 
Preprint IHEP 96--68, Protvino, hep-th/9609031.

\bibitem{Dub96}
Dubrovin, B.: Geometry of $2D$ topological field theories.  In:
Integrable Systems and Quantum Groups. Francaviglia, M., Greco S.
(eds.) Lecture Notes in Mathematics vol. {\bf 1620}. Berlin: Springer
1996, p.~120.

\bibitem{Hit96}
Hitchin, N.: Frobenius manifolds (Notes by D. Calderbank). Preprint,
1996. 

\bibitem{Man96}
Manin, Yu. I.: Frobenius manifolds, quantum cohomology, and moduli
spaces (Chapters I, II, III). Preprint Max-Planck-Institut f\"ur
Mathematik, Bonn, 1996.

\bibitem{LPSS95}
L\"u, H., Pope, C. N., Sezgin, E., Stelle, K. S.: Stainless super $p$-branes.
Nucl. Phys. {\bf B456}, 669--698 (1995).

\bibitem{Dub90}
Dubrovin, B. A.: Differential geometry of strongly integrable systems
of hydrodynamic type. Funct. Anal. and its Appl. {\bf 24(4)},
280--285 (1990).

\bibitem{ZSh74}
Zakharov, V. E., Shabat, A. B.: A scheme for integrating the
nonlinear equations of mathematical physics by the method of the
inverse scattering problem. Funct. Anal. and its Appl. {\bf 8(3)},
226--235 (1974); Integration of nonlinear equations of mathematical
physics by the method of inverse scattering. II. Funct. Anal. and its
Appl. {\bf 13(3)}, 166--174 (1979).

\bibitem{Vil69}
Vilenkin, N. Ya.: Fonctions sp\'eciales et th\'eorie de la
repr\'esentation des groupes, Paris: Dunod 1969.

\end{thebibliography}
\end{document}